\documentclass[grl]{agu2001}
\usepackage{graphicx}
\usepackage{rotating}
\usepackage{agu-ps}
\usepackage{amssymb}

\authorrunninghead{KR\"UGER ET AL.}
\titlerunninghead{JOVIAN DUST STREAMS: MONITOR OF IO'S PLUME ACTIVITY}

\journalid{} \articleid{}{} \paperid{\today}
\cpright{AGU}{2003}

\received{} \revised{}
\accepted{}\published{}

\authoraddr{H. Kr\"uger, S. Kempf, R. Srama, G. Moragas-Klostermeyer, 
R. Moissl, E. Gr\"un, 
MPI Kernphysik, Postfach 103980, 69029 Heidelberg, Germany
(E-Mail: Harald.Krueger@mpi-hd.mpg.de)}
\authoraddr{P. Geissler, LPL, U. of Arizona, 
Tucson AZ 85721, USA}
\authoraddr{M. Hor{\'a}nyi, LASP, 
U. of Colorado, Boulder, CO 80309, USA}
\authoraddr{A. L. Graps, IFSI, INAF, Via del Fosso del Cavaliere 100, 
00133 Roma, Italy}
\authoraddr{T. V. Johnson, JPL, %California Institute of Technology,
4800 Oak Grove Dr. Pasadena, CA 91109, USA}
\begin{document}

\title{Jovian Dust Streams: A monitor of Io's volcanic plume activity} 

\author{Harald Kr\"uger\altaffilmark{1},
        Paul Geissler\altaffilmark{2},
        Mih{\'a}{\l}y Hor{\'a}nyi\altaffilmark{3},
        Amara L. Graps\altaffilmark{1,4},
        Sascha Kempf\altaffilmark{1},
        Ralf Srama\altaffilmark{1},
        Georg Moragas-Klostermeyer\altaffilmark{1}, 
        Richard Moissl\altaffilmark{1},
        Torrence V. Johnson\altaffilmark{5},
        Eberhard Gr\"un\altaffilmark{1,6}
}

\altaffiltext{1}{Max-Planck-Institut f\"ur Kernphysik, Heidelberg, Germany}

\altaffiltext{2}{Lunar and Planetary Laboratory, Univ. of Arizona, Tucson, USA}

\altaffiltext{3}{Laboratory for Atmospheric and Space Physics and Department of Physics,
Univ. of Colorado, Boulder, USA}

\altaffiltext{4}{Inst. di Fisica dello Spazio Interplanetario, INAF, 
Roma, Italy}

\altaffiltext{5}{Jet Propulsion Laboratory, Pasadena, California, USA}

\altaffiltext{6}{Hawaii Inst. of Geophysics and Planetology,
Univ. of Hawaii, Honolulu, USA}

\begin{abstract}
Streams of high speed dust particles 
originate from Jupiter's
innermost Galilean moon Io. After release from Io,
the particles collect electric charges in the Io plasma torus, 
gain energy from the co-rotating electric field of Jupiter's 
magnetosphere, and leave the Jovian system into interplanetary 
space with escape speeds over 
$\rm 200\,km\,s^{-1}$. Galileo, which was the first orbiter
spacecraft of Jupiter, has continuously monitored the dust
streams during 34 revolutions about the planet
between 1996 and 2002. The 
observed dust fluxes exhibit large orbit-to-orbit variability
due to systematic and stochastic changes. After removal of the
systematic variations, the total dust emission rate of Io
has been calculated.
It varies between $10^{-3}$ and $\mathrm{10}\, \rm kg\, s^{-1}$,  
and is typically in the range of $0.1$ to $\rm 1\,kg\,s^{-1}$. 
We compare the dust emission rate with other markers of 
volcanic activity on Io
like large-area surface changes caused by volcanic deposits and 
sightings of volcanic plumes.
\end{abstract}

\begin{article}

\section{Introduction}

Io is the most volcanically active body in 
the Solar System. 
When the most dramatic signs of Io's volcanism -- its energetic 
volcanic plumes -- were discovered with the Voyager spacecraft in
1979, it was suggested that tiny dust grains entrained in the plumes
might be ejected into circumjovian space by electromagnetic forces 
\citep{johnson1980,morfill1980b}.

First observational evidence for this mechanism came with the
discovery of streams of dust particles with the dust detector on-board the 
Ulysses spacecraft in 1992 \citep{gruen1993}. 
These grains were detected more than 200 million kilometers away from Jupiter 
and they could be tracked to the Jovian 
system \citep{zook1996}. Derived grain radii were $\rm \approx 10\,nm$ 
and their speeds exceeded $\rm 200\,km\,s^{-1}$. 
The grains are accelerated to such high speeds by the
corotational electric field of Jupiter's magnetosphere 
\citep{horanyi1993a}.
Galileo dust measurements obtained within the Jovian
magnetosphere showed that the stream particles are strongly
coupled to the Jovian magnetic field \citep{gruen1998}, and 
%that the intensity of the dust fluxes is modulated by both Io's orbital 
%period and Jupiter's rotation period \citep{graps2000a}. 
Io was shown to be the ultimate source for the majority of the 
particles rather than other potential sources \citep{graps2000a}.

Voyager, Galileo and Cassini imaging observations have detected at least
17 volcanic centers with related plumes \citep{mcewen2003}.
Most of the plumes were sensed through the scattering of sunlight by
dust particles entrained within the plumes.
Ring-shaped surface deposits suggest that other plumes have been
recently active as well.
%At least two major types of plumes can be distinguished: large, faint 
%plumes, with short-lived or intermittent 
%activity (Pele-type) or small, bright, long-lived ones (Prometheus-type). 
%Prometheus, the archetype of the second category, is Io's
%most persistently active plume. 
Pele, one of the most powerful plumes on Io, has been observed 
at altitudes up to 460\,km \citep{spencer1997}. 

Since 1996 Galileo has made 34 orbits about Jupiter and provided a 
unique long-term record of the dust flux ejected from Io. In
particular, as the plumes are the most plausible sources
of the dust particles, the dust measurements monitor plume 
activity. The dust data show a large orbit-to-orbit variation 
due to both systematic and stochastic changes.
Systematic effects include Io's orbital motion, changes in
the geometry of Galileo's orbit and in the magnetic field configuration
due to the rotation of the planet. Stochastic
variations include fluctuations of Io's volcanic activity and the 
deformation of the outer magnetosphere in response to the variable 
solar wind conditions. By combining the entire Galileo dust data set, 
the variability due to stochastic processes
could be removed and a strong flux variation with Jovian 
local time showed up \citep{krueger2003a}, confirming earlier
predictions \citep{horanyi1997}.

In this paper we use the entire Galileo data set and examine the 
orbit-to-orbit variability of the dust emission pattern. We
derive the dust emission rate of Io and its variations during
the Galileo mission and compare it with
other markers of volcanic activity on Io.

\section{Galileo dust measurements at Jupiter}

The Galileo Dust Detector System (DDS), like its twin on-board 
Ulysses, is a multi-coincidence impact ionization 
detector which measures submicron- and micron-sized dust particles
\citep{gruen1992a,gruen1995b}.
DDS is mounted on the spinning section of Galileo
and has a 140\deg\ wide field of view. 
During one spin revolution of the spacecraft, the
detector scans the entire anti-Earth hemisphere. 
Jovian streams particle identification has been
previously described in the literature 
\citep{gruen1995b,gruen1998,krueger1999c,krueger2001a}. 

\begin{figure}%[h]
\vspace{-5mm}
\begin{turn}{90}
\includegraphics[width=0.74\hsize]{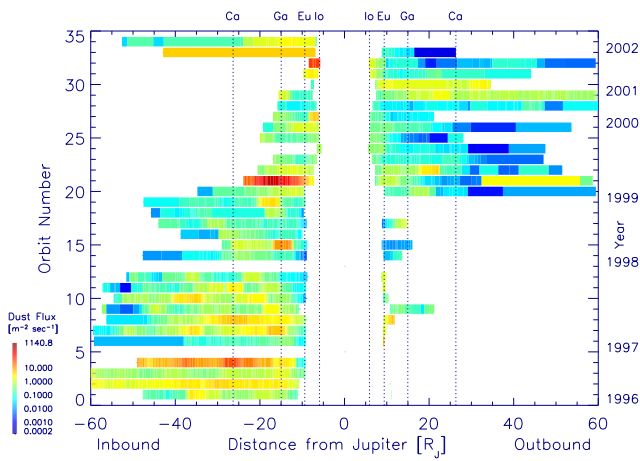}
\end{turn}
\vspace{-5mm}
    \caption{
Dust fluxes measured along 34 Galileo orbits about Jupiter 
vs. jovicentric distance. Years are indicated on the right.
Fluxes were corrected for DDS aging. 
%A DDS sensitive area of 
%$\rm 20\,cm^{2}$ is indicated by a red solid line. 
The orbits of
the Galilean moons are indicated by vertical dotted lines. 
%Thin horizontal lines show intervals when RTS data were collected.
Dust fluxes have been smoothed with a 10~h boxcar average.
%Flux values for less than $\rm 20\,cm^{2}$ sensor area were ignored.
%Due to the changing detection
%geometry of DDS during the Galileo mission (see text), the dust 
%streams were detected in varying spatial regions.
        }
    \label{fluxplot}
\end{figure}

The Galileo trajectory remained in 
Jupiter's equatorial plane within a few degrees.
As the line of apsides of the spacecraft trajectory stayed constant
in an inertial frame, the orbital motion of Jupiter 
about the Sun lead to a 360\deg\ shift 
with respect to the Sun-Jupiter line
within one Jovian year (11.9 Earth years). 
The shift of the apsidal line resulted in a 
a gradual change of the detectability of the jovian stream
particles
during the Galileo Jupiter mission. Early in the mission, 
stream particles could be mostly detected in the inner Jovian 
system, i.e. within $\sim 50\,\rm R_J$ from the planet, with a varying
detection geometry from orbit to orbit 
(Jupiter radius $\rm R_J = 71,492\,km$). Only after mid-2000
was the detection geometry significantly different so that the streams 
could be detected during almost the entire Galileo orbit.
Highly time resolved real-time science 
(RTS) data were received from the dust instrument during most of the time 
periods when the dust streams were detectable. Dust fluxes measured 
within $\rm 60\,R_J$ from 
Jupiter are shown in Fig.~\ref{fluxplot}.
In these time periods and this spatial region the flux 
varied by almost six orders of magnitude 
($\approx 10^{-3}$ -- $10^{3} \, \rm m^{-2} \, s^{-1}$). 

To determine the dust flux, we simultaneously multiplied the rate
of dust stream impacts and the effective sensitive area of the 
detector averaged over one spacecraft spin 
revolution, as Galileo moved about Jupiter. To calculate 
the approach direction, we used a ``standard'' trajectory of a 10~nm
radius particle \citep{gruen1998}. Particles of this size 
best explained observed times of 
of onset, a $180^{\circ}$ shift, and cessation of 
stream particle impacts which were recognised during many
Galileo orbits.
Flux values for sensitive areas below $20\, \rm cm^{2}$ 
were ignored because the derived fluxes are rather uncertain.

The sensitivity of DDS has changed over time due to radiation-related 
aging effects in the instrument electronics. 
This degradation has been revealed 
from the electronics' response to simulated charge 
signals generated by a test pulse generator which is part of DDS.
The change of the instrument response is consistent 
with a gradual shift of the charge signals measured for the 
stream particles.
The drop in sensitivity
has been a factor of 5 since Galileo's early Jupiter mission in 1996.
By assuming a size distribution of the particles, a time-dependent 
correction
for the flux can be calculated and applied to the data:
taking the impact charge distribution measured for stream particles
in 1996 
%(\cite{krueger2001a}, 
[\markcite{{\it Kr\"uger et al.}, 2001}] when DDS
still had its nominal sensitivity, the 
flux correction factor is one in 1996 and gradually rises up to 25 
in 2002.

Significant variations in the dust flux 
orbit-to-orbit
are obvious in Fig.~\ref{fluxplot}. In addition, 
for a source with a constant dust ejection rate 
and assuming continuity, one expects the dust flux to drop with the 
inverse square of the source distance. This is only rarely the case.
Particularly high fluxes were measured during orbits E4 
and C21.

\section{Io's dust emission rate}

How significant is Io as a source of cosmic dust? 
A simple calculation yields the total dust emission
rate of Io. Ulysses, Galileo and Cassini dust measurements in 
interplanetary space have recorded the dust streams in the 
ecliptic plane.
In addition, Ulysses also measured the streams at $-35^{\circ}$ 
jovigraphic latitude in 1992, and 
very recently, in July 2003, detected the dust 
streams even at $+55^{\circ}$ latitude.
This is consistent with models for the
electromagnetically dominated dynamics of the grains within the
Jovian magnetosphere \citep{horanyi1993a}. 
The long record of 
Galileo dust stream measurements from 1996 to 2002 has shown that the 
Io torus acts as a 'smeared out (ring) source` rather than a 
sharply localized point source \citep{krueger2003a}.
Given the spread of Io dust along and away from Jupiter's
equatorial plane, we assume a wedge-shaped emission pattern 
of dust originating at Jupiter, with a wedge opening angle 
$\alpha = 35^{\circ}$ 
(this angle is implied by our Ulysses
measurements).
Furthermore, we assume 
a constant particle speed. Dynamical modeling shows that
most of the
particle acceleration occurs in the region within Europa's orbit
($\rm \lesssim 10\,R_J$). By ignoring flux measurements from within
$13\, \rm R_J$ from Jupiter the acceleration further out is 
negligible for our calculation. Finally, a 'dilution` of the 
dust with $d^{-2}$, $d$ being 
distance from Jupiter, is expected from continuity.
Then, the 
total mass of dust ejected from Io [$\rm kg\,s^{-1}$] 
is:
\begin{equation}
W =
 \frac{16}{3} \pi^2 s^3\, F_d \, \rho \, d^2 f(\Lambda) 
\, \tan \alpha . \label{dustprod}
\end{equation}
The dust flux $F_d$ at distance $d$ is shown in Fig.~\ref{fluxplot}. 
For the grain density 
we assume $\rho = 1.5\,{\mathrm g\,\mathrm{cm}^{-3}}$. The grain radius 
is $s = 10\,{\mathrm{nm}}$ \citep{zook1996,gruen1998}. 
A variation of the dust flux with Jovian local time, $\Lambda$, 
\citep{krueger2003a}, is
described by $f(\Lambda)$ and is 
typically below a factor of 5 and, hence, much smaller than the
flux variations seen in the data. 

The assumed grain size of 10\,nm is an effective radius 
as implied by our DDS measurements and is close to the detection threshold
of the instrument. The abundance of bigger grains is very strongly
decreasing. Much smaller particles, however,  are likely to be generated
by Io, but they are not likely to be able to leave the Io torus because 
of their electromagnetic
interaction with the magnetosphere. Therefore, dust ejection rates
derived in this paper should be treated as lower limits. Given our simple
assumptions made in Eqn.~\ref{dustprod} the absolute dust emission rates
are accurate to an order of magnitude, at most. The accuracy of the 
relative variations with time, however, should be more accurate.

In principle, Eqn.~\ref{dustprod} allows us to calculate dust emission
rates from measurements at any distance from Jupiter. 
However, the dust dynamics are only reasonably well understood in the
inner Jovian magnetosphere, and, with our assumed
simple wedge-shaped emission pattern, dust measurements very far 
from Jupiter may erroneously give very large emission rates (Sect. 4). 
We therefore consider emission rates for $13 < d < 30\, \rm R_J$ 
as the most reliable. 
Galileo 
typically spent only several days in this spatial region 
during each orbit
so that emission rates were derived from this short time interval.

The dust emission rates derived from the DDS measurements
are shown in Fig.~\ref{iodustprod}. The emission rate 
for $13 < d < 30\, \rm R_J$ varied between 
$10^{-3}$ and $\mathrm{10}\, \rm kg\, s^{-1}$,  
with a typical average of $0.1$ to $\rm 1\,kg\,s^{-1}$. This 
value is consistent with theoretical predictions assuming
effective electrostatic charging in the upper
regions of the plumes \citep{ip1996}.
Compared with 
$\sim 10^3\,\rm kg\,s^{-1}$ of plasma ejected from Io into the
torus, the dust amounts to only 0.01 to 0.1\% of the total mass released.
These numbers indicate that Io is also a minor 
source of interplanetary dust compared with comets and main belt
asteroids 
($\sim 10^4\,\rm kg\,s^{-1}$).
Io, however, turns out to be a major dust source for the Jovian system
itself. The total mass of dust produced by Io as 10 nm 
particles is comparable to the mass of dust ejected as micrometer-sized
particles due to hypervelocity impacts of interplanetary micrometeoroids 
by the other Galilean satellites 
which have no volcanic activity \citep{krueger2003b}.

In 2000 (orbit G28) Galileo left the Jovian magnetosphere for the 
first time since 1995 to a jovicentric distance of 
$\sim 280 \rm R_J$ (0.13~AU). In this time period 
a high dust flux of up to 
$\rm \sim 10\,m^{-2}\, sec^{-1}$
was measured for about two months. 
%comparable with the 
%rates detected both in interplanetary space and close to Jupiter during 
%Galileo's earlier orbital mission. 
%Assuming again $\alpha = 35^{\circ}$, 
This leads
to a dust emission rate of $\sim 100\,\rm kg\, s^{-1}$ (dashed line in 
Fig.~\ref{iodustprod}). 
%Whether it is really due to strong dust
%emission of Io, or to compression of the Jovian magnetosphere, or both,
%is presently unknown. The interaction of the Jovian magnetosphere with 
%the interplanetary magnetic field may have led to focussing of the
%dust grains. 
Similarly high fluxes
were also recorded with the 
Cassini dust instrument at $\sim 0.3$AU from Jupiter when the spacecraft was 
approaching the planet in September 2000.

Estimates of the Io dust emission rate from the Galileo and Ulysses 
dust measurements in interplanetary space out to 1\,AU from Jupiter -- 
assuming again that the dust is uniformly distributed into a wedge of 
$35^{\circ}$ -- lead to unrealistically
high dust emission rates of more than $10^7 \rm kg\, s^{-1}$. It 
clearly shows 
that our simple picture cannot be extrapolated into interplanetary 
space. 
%Cassini dust measurements imply that the particle dynamics there 
%are presently not understood: dust stream 
%particles have even been detected from the anti-Jovian direction which 
%cannot be explained with existing models. In February 2004 Ulysses
%will approach Jupiter to 0.8~AU at high jovigraphic 
%latitudes and will thus be able to
%measure the Jovian dust streams in interplanetary space again. This 
%may give new insights into the particle dynamics in interplanetary
%space.

%One might argue that the assumed wedge opening angle  $\alpha = 35^{\circ}$
%is too large and that a wedge angle constrained by the $10^{\circ}$ 
%tilt of Jupiter's magnetic field axis is more realistic \citep{johnson1980}. 
%Equation~\ref{dustprod} shows that the dust emission rate derived 
%for $\alpha = 10^{\circ}$ is smaller by a factor of 4.

It should be noted that the maximum and minimum dust emission values in C21 
derived from $ 13 < d < \rm 30\,R_J$ are four orders of magnitude apart
(Fig.~\ref{iodustprod}). The dust emission varied by such a large amount 
within only four days (Days 182 to 185 in 1999).

\section{Dust emission versus volcanic activity}

\begin{figure}[t]
\vspace{-4mm}
\hspace{-8mm}
\includegraphics[width=1.09\hsize]{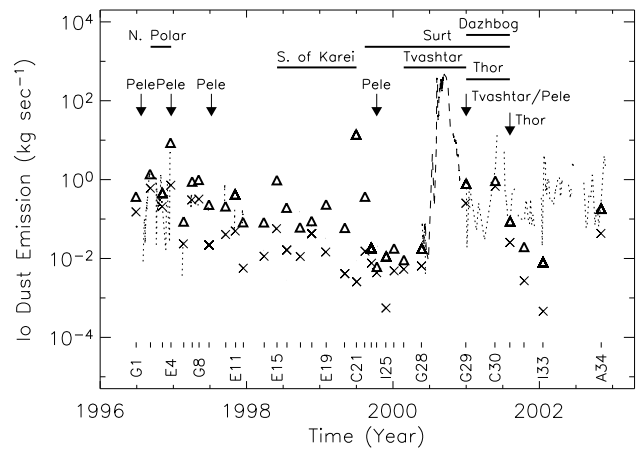}
\vspace{-7mm}
    \caption{
Dust emission rate of Io calculated with Eqn.~\ref{dustprod}. Triangles
and crosses denote the maxima and minima derived from measurements in the 
distance range $ 13 < d < 30\rm \,R_J$, respectively. The dashed line 
is for the G28 orbit in the range $ 30 < d < \rm 280\,R_J$, dotted lines 
show the remaining orbits with 
$ 30 < d < \rm 400\,R_J$. Thick horizontal bars indicate 
periods when large-area surface changes occurred on Io
\citep{geissler2003}, arrows indicate individual plume sightings.
Note that South of Karei probably erupted just 
before C21, and that Surt's eruption probably took place between G29 and C30 
(see text). 
Galileo perijove passages (vertical dashes) and orbit labels are indicated 
at the bottom. No dust stream measurements were collected during Galileo 
orbits 5 and 13.
        }
    \label{iodustprod}
\end{figure}

Exceptionally high dust emission rates occurred during
orbits E4 and C21, G28, and to a lesser extent also during 
G29 and C30 (Fig.~\ref{iodustprod}). Hence, it is of particular 
interest whether these peaks 
%in the dust emission
can be related to specific plume sightings or other observations of
volcanic activity on Io. 

To correlate dust measurements collected far away from Io with 
imaging observations one must  
know how long the particles need to travel from the source
to the Galileo spacecraft. After ejection from Io, the particles 
typically need several hours up to one day to collect sufficient
charge and get ejected from the torus \citep{horanyi1997}. After 
release from the torus, they need several hours to travel 
to a jovicentric distance of $\rm 30\,R_J$ and one day to travel to
$\rm 280\,R_J$, respectively. 
Therefore, after ejection from a plume, the particles reach the
spacecraft within one to two days, which, for our analysis, is a 
negligible time delay.

A correlation of the dust emission with the activity of the most 
energetic plumes seems most promising. Only these plumes are expected to 
accelerate the grains to high altitudes so that they can 
collect sufficient charge from the ambient plasma to overcome
the satellite gravity \citep{johnson1980,ip1996}. 
A correlation of our in-situ dust measurements with either Galileo
or Earth-based plume imaging observations turned out to be very difficult. 
Because of the incomplete temporal coverage, only a very incomplete record 
of plume activity is available from direct sightings \citep{mcewen1998,keszthelyi2001}. 
This is further complicated because
plume activity sometimes changes on timescales of days to weeks. 

A more complete history of Io's explosive eruptions has been derived from
a study of the surface changes they produced \citep{geissler2003}.
These surface changes are caused by plume eruptions that can be divided
into two categories. Smaller plumes produce near-circular rings typically
150 to 200 km in radius that are white or yellow in color unless
contaminated with silicates, and frequently coat their surroundings with
frosts of fine-grained $\rm SO_2$. Giant plumes are much rarer, limited to a half
dozen examples, and produce oval, orange or red, sulfur-rich rings with
radii in the range from 500 to 550 km. Because they are much more
energetic than the smaller plumes and eject dust particles at speeds
approaching Io's escape velocity, the giant plumes probably contribute
most to the escape of dust from Io.

Giant plume eruptions occurred at the North Pole of Io (North Polar Ring)
during the time interval between Galileo's G2 and E4 orbits in 1996, in
the region South of Karei between orbits E15 (1998) and C21 (1999), at
Surt between C22 (1999) and I31 (2001) and at Dazhbog between G29 and C30
(2001, Fig.~\ref{iodustprod}). 
An enormous red ring appeared around Tvashtar which most
likely erupted between orbits I27 and G29 in 2000. Pele's deposits were
gradually replaced after Pillan's eruption obscured parts of Pele's red
ring during orbit C9. Plume sightings and observations of thermal emission
provide closer constraints on the dates of some of these eruptions. Surt's
eruption probably took place in February 2001 (between G29 and C30), based
on high-temperature lavas spotted in ground-based observations 
\citep{marchis2002}.
South of Karei probably erupted just before orbit C21, based
on thermal emission detected by Galileo \citep{keszthelyi2001} and
ground-based observations \citep{howell2001}. The North Polar ring may
have been associated with a large thermal event seen from Earth in October
1996 \citep{stansberry1997} but its location is too uncertain to be
sure. In addition, a 400km high plume was imaged at Tvashtar with the
Cassini spacecraft during its Jupiter fly-by in early January 2001 
\citep[Galileo orbit G29]{porco2003}.
Many smaller eruptions also took place
during the Galileo Mission, including a dramatic eruption at Pillan during
orbit C9 \citep{mcewen1998} and a large plume at Thor seen by Galileo
during orbit I31 \citep{mcewen2001}, but their deposits were much less
extensive than those of the giant plumes and they are unlikely to have
contributed substantially to the flux of dust escaping Io.

In many cases, the eruptions of giant plumes are in agreement 
with the time periods when our in-situ dust-measurements showed episodes 
of elevated dust emissions.
These coincidences
indicate that the observed surface changes \citep{geissler2003} were 
indeed due to very powerful plume eruptions, and
the extent of the changes implies that these eruptions 
must have been among the most powerful during the 
Galileo mission. 
The lack of plume sightings and/or recognition of large area
surface changes in some time intervals when strong dust emission was 
observed, 
may point to volcanic eruptions on Io which were only 
recognised because of increased  dust emission. Here, the dust
measurements may eventually serve as a monitor of Io's volcanic
activity.

If the high dust emission rates deduced from the dust measurements
are indeed connected to the largest explosive 
eruptions of volcanic plumes, which plumes caused the dust emission at a 
lower level during the remaining periods? Surface changes have been 
detected at many other locations, and plumes have also been seen at other
times. In particular, the Prometheus plume may be a continuous
supplier of dust on a low level if the particles entrained in it 
are ejected with sufficient speed. At least, the imaging observations 
imply that is was continuously active, its prominent 50 
to 150~km high dusty plume  was visible
in every image taken of it by both Voyagers and Galileo \citep{geissler2003}.

Another candidate is Pele, one of the most powerful plumes and 
the most steady high-temperature volcanic center
on Io. Surface changes at the Pele site were detected frequently
during the Galileo mission whereas detections of the Pele plume 
are relatively rare. During the Galileo Jupiter mission 
the Pele plume was observed in July 1996 (G1), December 1996 (E4) possibly
in July 1997 (C9), 
October 1999 (I24)
and by the Cassini
spacecraft in January 2001 (G29) \citep{mcewen1998,porco2003}. 
On the other hand, it was absent in June 1996, February 1997, June 1997 and 
July 1999. Hence, two detections of the Pele plume are also coincident with
our measurements of high dust fluxes in E4 and G29.
The low dust flux detected in E6 may be explained by the absence 
of the Pele plume.

Particle radii derived from our in-situ measurements are $\sim \rm 10\,nm$
\citep{zook1996,gruen1998}. Theoretical considerations imply that particles of
this size and below may be able to escape Io \citep{johnson1980,ip1996}.
The Pele plume is rich in short-chain sulfur ($\rm S_2$) as well as 
$\rm SO_2$ \citep{spencer2000}
%Although it has been suggested that Pele may be a pure gas
%plume, plume observations can also be interpreted as being due to 
and contains 
very fine ($ \rm \leq 80\,nm$) particulates \citep{spencer1997}. This is 
consistent with Voyager observations of the Loki plume \citep{collins1981}.

The dust emission rates derived from our Galileo DDS measurements
reflect lower limits to
the amount of dust generated in the eruptions on Io and ejected 
into circumjovian space. The ejected dust, however, represents only a 
small fraction of the total amount of solids erupted. By far the
largest amount of material is deposited on Io's surface. Our dust 
measurements may lead to improved plume models and, together with Galileo 
imaging observations, to better constraints of the total amount of material 
deposited on the surface.

\begin{acknowledgments}
We are grateful to John Spencer for providing us with Io thermal emission 
data.
We thank the Galileo project at JPL for effective and successful 
mission operations.
This research has been supported by the German Bundesministerium f\"ur Bildung 
und Forschung through DLR
(grant 50 QJ 9503 3). M.H. has been supported by NASA.
\end{acknowledgments}

%\bibliography{/home/krueger/tex/bib/pape,/home/krueger/tex/bib/references}

\end{article}

%\pagebreak

%{\Large \bf Figure Captions}

\bigskip
\bigskip

%\begin{figure}[h]
%%\hspace{1.5cm}
%%\includegraphics[width=0.6\hsize]{fig2}
%\includegraphics[width=0.6\hsize]{fig2.eps}
%%\epsfxsize=0.6\hsize
%%\epsfbox{../bilder/fig2.eps}
%    \caption{
%Ejection geometry of Io dust into circumjovian space. 
%        }
%    \label{sketch}
%\end{figure}

\end{document}